\documentclass{article}
\topskip=\baselineskip
\let\dl=\delta
\let\ep=\epsilon
\let\et=\eta
\let\th=\theta
\let\ka=\kappa
\let\lm=\lambda
\let\om=\omega

\let\pa=\partial
\let\e=\emph

\let\ct=\cite

\let\bv=\mathbf

\let\dt=\cdot
\let\del=\nabla

\let\rta=\rightarrow
\let\dy=\displaystyle
\let\ty=\textstyle

\let\hl=\hfill
\newcommand{\m}{\mbox}
\newcommand{\ol}[1]{\makebox[\textwidth][s]{#1}}

\newcommand{\eqdf}{\stackrel{\mathrm{def}}{=}}
\newcommand{\hf}{\ensuremath{{\scriptstyle\frac{1}{2}}}}

\newcommand{\be}{\begin{equation}}
\newcommand{\ee}{\end{equation}}
\newcommand{\dd}[3]{\\ \m{}\\ \ol{\m{#1}\hl\m{${\dy #2}$}\hl\m{#3}}\\ \m{}\\}
\newcommand{\re}[2]{\dd{}{#1}{(#2)}}

\newcommand{\ba}{\begin{array}}
\newcommand{\ea}{\end{array}}
\newcommand{\bea}{\begin{eqnarray}}
\newcommand{\eea}{\end{eqnarray}}
\newcommand{\beas}{\begin{eqnarray*}}
\newcommand{\eeas}{\end{eqnarray*}}

\newcommand{\bx}{\pa_\lm\pa^\lm}

\newcommand{\vE}{\bv{E}}

\newcommand{\vj}{\bv{j}}
\newcommand{\vA}{\bv{A}}

\newcommand{\vk}{\bv{k}}
\newcommand{\vr}{\bv{r}}

\title{Covariant isolation from an Abelian gauge field \\
       of its nondynamical potential, which, when fed \\
       back, can transform into a ``confining Yukawa''}
\author{Steven Kenneth Kauffmann \\
        American Physical Society Senior Life Member}
\date{43 Bedok Road \\
      {\#}01-11 \\
      Country Park Condominium \\
      Singapore 469564 \\
      Handphone: +65 9370 6583 \\
      \m{} \\
      and \\
      \m{} \\
      Unit 802, Reflection on the Sea \\
      120 Marine Parade \\
      Coolangatta QLD 4225 \\
      Australia \\
      Tel/FAX: +61 7 5536 7235 \\
      Mobile:  +61 4 0567 9058 \\
      \m{} \\
      Email: SKKauffmann@gmail.com}
\begin{document}
\maketitle
\begin{abstract}
For Abelian gauge theory a properly relativistic gauge is developed by
supplementing the Lorentz condition with \e{causal} determination of the time
component of the four-vector potential by retarded Coulomb transformation of
the charge density.  This causal Lorentz gauge \e{agrees} with the Coulomb
gauge for \e{static} charge densities, but allows the four-vector potential to
have a \e{longitudinal} component that is determined by the time derivative of
the four-vector potential's time component.  Just as in Coulomb gauge, the two
\e{transverse} components of the four-vector potential are its sole
\e{dynamical} part.  The four-vector potential in this gauge covariantly
\e{separates} into a dynamical transverse four-vector potential and a
nondynamical timelike/longitudinal four-vector potential, where \e{each} of
these two satisfies the Lorentz condition.  In fact, analogous partition of the
conserved four-current shows \e{each} to satisfy a Lorentz-condition
Maxwell-equation system with its \e{own} conserved four-current.  Because of
this complete separation, either of these four-vector potentials can be
tinkered with \e{without affecting its counterpart}.  Since it satisfies the
Lorentz condition, the nondynamical four-vector potential times a constant with
dimension of inverse length squared is itself a conserved four-current, and so
can be fed back into its own source current, which transforms its time
component into an extended Yukawa, with \e{both} exponentially decaying \e{and}
exponentially \e{growing} components.  The latter might be the mechanism of
quark-gluon confinement: in non-Abelian color gauge theory the Yukawa mixture
ratio ought to be tied to color, with palpable consequences for ``colorful''
hot quark-gluon plasmas.
\end{abstract}

\subsection*{Introduction}

Gauge theories have a quintessential \e{dual nature}.  They
\e{simultaneously} encompass \e{dynamical} transverse waves
that travel at the speed of light, which phenomenon
can be \e{independently} quantized, and \e{nondynamical}
potential/force fields whose \e{source} is appropriately
``charged'' matter, and which, in turn, \e{affect the behavior}
of such matter.  The \e{nondynamical pure potentials/forces},
which are \e{part} of what gauge theories encompass, are \e{not}
subject to \e{independent} quantization: they \e{merely conveniently
isolate and abstract} a certain \e{intermediate mathematical aspect}
of physical behavior that is \e{inherent} to the ``charged'' matter
\e{itself}---useful mathematical abstractions, however suggestive or
extremely convenient, are \e{still} not \e{physical degrees of free%
dom} that can be \e{independently} quantized; any \e{quantum} charac%
teristics of such nondynamical pure potential/force fields are merely
the \e{secondary consequences} of the \e{quantum character} of their
``charged'' matter \e{source}.

In view of the \e{dual} nature of gauge theories, their typically
\e{seamless}-appearing \e{mathematical} formulations present a
conundrum and challenge to the theoretical physicist.  Part of the
formal smoothness of gauge theories is attributable to the physics
which they describe: the selfsame ``charged'' matter which inherently
interacts with \e{itself} in a manner that can be conveniently
mathematically analyzed using the intermediate constructs of
source-determined pure potential/force fields, \e{also} emits, absorbs
and scatters the \e{dynamical} transverse waves/radiation/massless
quanta---whose \e{interactions} with that ``charged'' matter are made
\e{part and parcel} of the \e{physics} treated by that \e{same} gauge
theory.  This naturally promotes the \e{formal similarity} of fields
that \e{do} have \e{dynamical content} to fields which are merely
extremely convenient \e{intermediate mathematical constructs} for the
analysis of ``charged'' matter's \e{intrinsic} interaction.  Another
technical aspect of gauge theories which can contribute to such a lack
of formal distinction between actual \e{dynamical} fields and \e{non%
dynamical} ``intermediate mathematical construct'' pure \e{potential}
fields is the gauge invariance ambiguity of those theories: the
\e{physically nonexistent degree of freedom} often serves to \e{enhance}
misleading \e{surface} manifestation of \e{physically nonexistent sym%
metry}.

There are a number of reasons to seek to, in a formally neat and relativ%
istically covariant fashion, ``pull apart'' gauge theory into two physic%
ally natural parts, one encompassing the \e{purely dynamical transverse
gauge fields} and the other the \e{purely nondynamical gauge potential
fields}.  Obviously, pushing the nondynamical potential fields out of the
way in a covariant fashion might conceivably be a boon to covariant
quantization of the dynamical transverse gauge fields.  More intriguingly,
having a clean such separation in hand raises the theoretical possibility
of tinkering with one of these parts without affecting the other, or perhaps
even trying to discard one of the parts altogether.  Certainly the issue of
quark confinement lends itself to thoughts of somehow relativistically
drastically reshaping ``gluon'' gauge \e{potential fields} without upending
altogether the notion of the gluon as a gauge particle.  The last half of
this paper tentatively explores just this matter, albeit only in the
theoretically insufficient context of purely Abelian gauge theory.

As a first step along this route, this paper deals only with the simplest
case, namely Abelian gauge theory.  Hopefully there will be other
researchers who will be inspired to look at the more involved non-Abelian
cases.  The approach to carrying out covariant cleavage of Abelian gauge
theory into two natural constituents will lean heavily on the \e{identi%
fication of a sensible potential field}.  Since the potential field is an
intermediate mathematical construct for facilitating the understanding
of the \e{intrinsic} interaction of charged matter, it should be
\e{entirely the creature} of its charged matter \e{source}; in other
words, with the charged matter \e{source} in hand, the potential field
should \e{follow uniquely}, and it should, of course, \e{vanish alto%
gether} in the \e{absence} of the charged matter \e{source}.  Since the
equations of Abelian gauge theory are \e{linear}, it is clear that the
potential field must be a \e{homogeneous functional} of its \e{source}.
Furthermore, the \e{kernel} of the \e{appropriate homogeneous functional}
is a \e{Green's function} of the linear equation that relates the poten%
tial field to its charged matter source.  To ensure that a \e{relativis%
tically sensible} Green's function is \e{available} for this purpose, we
must take care that our \e{choice of gauge} for the Abelian gauge theory
does \e{not} impose as the equation which links the potential field to
its charged matter source one that is \e{incompatible with special rela%
tivity}.  In other words, success of our envisioned project hinges on a
relativistically compatible \e{choice of gauge}.  It is to be noted that
the physics of the \e{dynamical} gauge waves/radiation/massless particles
in contrast does \e{not enslave} their transverse fields to a charged mat%
ter source; those fields can have a nonvanishing existence even in the
\e{complete absence} of charged matter, albeit they \e{interact} with
charged matter if it is present.  But their fields are \e{not determined}
by the charged matter, whereas the pure potential field \e{is} entirely so
determined, \e{even} in a \e{fully quantum context}.

At first blush it might appear that selection of the Coulomb gauge
of itself accomplishes the project just set out: at one stroke not
only is gauge ambiguity abolished, but also potential effects are
transparently assigned to the time component of the four-vector
potential, dynamical effects are equally transparently assigned to
the four-vector potential's spatial transverse components, and its
spatial longitudinal component is set to \e{zero} by fiat.  Every%
thing, apparently, is now clear-cut and simple.  The shock comes
when examining the full consequences for that time-component poten%
tial: it responds \e{instantaneously throughout all of space} to any
time variation of its charge-density source.  Thus the penny drops:
the marvelously simple and efficacious abolition of the longitudinal
component of the spatial vector potential, which \e{is} the Coulomb
gauge condition, spared no thought or concern for special relativity! 
Still, the Coulomb gauge's treatment of the \e{static limit} of the
charge density is absolute bedrock; any proposed alternative gauge
won't be worth its salt if it deviates from that.

The obvious alternative to the Coulomb gauge is the celebrated
Lorentz condition.  No slighting of relativity there, and an
immediate payoff with a marvelous automatic formal simplification
of the Maxwell equation system.  But hold on, doesn't that equation
system now look \e{too} symmetrical in the \e{components} of the 
four-vector potential?  Is there the \e{slightest indication}
of just \e{where} to seek the nondynamical potential versus the
dynamical radiation fields?  And what about the troubling
restricted gauge invariance ambiguity which still remains?

In fact, it happens that the Lorentz condition \e{itself}, with
\e{no reference} to the Maxwell equations, \e{completely deter%
mines} the the spatial longitudinal component of the four-vector
potential in terms of the time derivative of its time component.
Thus although the Lorentz condition \e{does not flatly abolish}
the longitudinal component of the vector potential as the Coulomb
gauge does, it \e{does completely subordinate it} to the time
derivative of the four-potential's time component, which has a
similar effect.  A puzzling feature of this consequence of the
Lorentz condition is that it is far from obvious how it could be
demonstrated by manifestly covariant operations on four-vectors.
The obvious approach is to note the precise mathematical analogy
between the Lorentz condition and the Coulomb-law Maxwell equation,
and then to recall that the latter determines the longitudinal
component of the electric field in terms of the charge density.
The determination of the longitudinal component of the spatial
vector potential in terms of the time derivative of its time
component is then worked out in strict analogy.

Though very helpful and enlightening, this consequence of the Lorentz
condition leaves a ``loose screw'' in its wake; the restricted
gauge invariance ambiguity that the Lorentz condition permits must
still be addressed.  Since the Lorentz condition ties the longitudinal
space component of the four-vector potential to its time component,
it is apparent that if the relation of the time component of the four-
vector potential to the \e{charge density} could have all the slack
\e{removed} from it (\e{without} offending special relativity, of
course), then that time component would be the full-fledged potential
field, and therefore the \e{dynamical} field would be \e{compelled} to
occupy \e{only} the two \e{transverse spatial} components of the four-%
vector potential, being that the four-vector potential's \e{longitudinal
spatial} component has \e{already} been \e{fully determined} by the
Lorentz condition to be \e{entirely} the creature of the time derivative
of what is now the \e{potential field}.

The key to removing the slack from the relation of the time component of
the four-vector potential to the charge density---which slack is a mani%
festation of the restricted gauge invariance ambiguity---of course lies
with the Green's functions of the light-speed wave equation \e{which con%
nects that time component to the charge density after imposition of the
Lorentz condition}.  The physically most appealing such light-speed wave-%
equation Green's function to select is clearly the celebrated traditional
causal \e{retarded} one, which makes perfect intuitive relativistic sense.
The result bears a considerable resemblance to the consequences of using
Coulomb gauge, except that the potential field is now tied to the charge
density in a relativistically causal \e{retarded} fashion rather than in
a relativistally offensive \e{instantaneous spatially uniform} fashion,
and the \e{time derivative} of that potential field now \e{fully deter%
mines} the \e{longitudinal} spatial component of the four-vector potential,
\e{instead} of its being decreed to \e{vanish under all circumstances}.
The upshot for the \e{dynamical} gauge field is \e{exactly the same} as it
is in Coulomb gauge: that field is described by the \e{two remaining trans%
verse} spatial components of the four-vector potential.  In the charge
density's static limit, the causal \e{retarded} potential field \e{as well}
comes out to be \e{static}, and therefore the \e{longitudinal} spatial
component of the four-vector potential \e{vanishes identically}.  This last
is, of course, the Coulomb gauge condition, so that in the charge density's
\e{static limit}, there is \e{no difference} from the Coulomb gauge
result.  Obviously this selection of the causal \e{retarded} Green's
function to specify a completely determined relation of the time
component of the four-vector potential to the charge density, in
\e{addition} to the imposition of the Lorentz condition, has \e{also}
determined a \e{gauge}, which we dub the causal Lorentz gauge.  This 
gauge is as close to the Coulomb gauge as one can get \e{without}
contravening the tenets of special relativity.

The causal Lorentz gauge permits the full four-vector potential to
be written as the sum of \e{two physically distinct} four-vector
potentials, the \e{first} one consists of only the \e{time and
spatial longitudinal components} of that full four-vector potential
in this gauge, while the \e{second} one consists of just the
\e{remaining two spatial transverse components} of that full four-%
vector potential in this gauge.  This separation is covariant
because \e{both} of these ``pulled apart'' four-vector potentials
\e{satisfy the Lorentz condition}.  Obviously the first, purely
timelike/longitudinal four-vector potential in this causal Lorentz
gauge is totally determined by the charge density, and is thus
entirely in the nature of a collection of potential fields, whereas
the second, remaining purely transverse four-vector potential in
this causal Lorentz gauge is purely dynamical.

Now the four-current conservation constraint condition is highly anal%
ogous to the Lorentz condition, and permits the four-current itelf to
be \e{similarly} partitioned into the sum of \e{two conserved four cur%
rents}, the first one encompassing the time component of the total four%
-current (i.e., the charge density) as well as the spatial longitudinal
component of the total four-current, which turns out to be \e{completely
determined by the charge density's time derivative}.  It is readily shown
that this first timelike/longitudinal four-current is \e{explicitly con%
served}.  The second four-current consists of just the two \e{remaining}
purely \e{transverse} components of the total four-current, and is readi%
ly shown to be \e{conserved as well}.  With this split up of the four-cur%
rent into a conserved timelike/longitudinal part plus a conserved purely
transverse part complementing the corresponding split up of the four-vector
potential into two highly analogous parts which each satisfy the Lorentz
condition, it turns out that Lorentz-condition light-speed wave-equation
versions of the Maxwell equations neatly apply to \e{both} of the split-up
four-vector potentials in causal Lorentz gauge.

We thus are now in a position to \e{deal entirely separately} with the
pure potential and the pure dynamical parts of the Abelian gauge theory
in causal Lorentz gauge.  Being that \e{adherence to the Lorentz condi%
tion is a pervasive feature of this gauge}, the four-vector potentials
\e{themselves} times a constant with the dimensions of inverse length
squared qualify as \e{conserved currents}.  In other words, in this
gauge one can envision \e{feeding back} the four-potentials as partial
contributors to their \e{own source current}.  To do this to the purely
transverse \e{dynamical} part of the four-vector potential seems tanta%
mount to giving the gauge particles \e{mass}, which is physically ques%
tionable in view of their \e{transverse} character: such particles with
\e{mass} are supposed to have \e{integer spin}, which is incompatible
with the \e{two} transverse spin states of such particles when they are
\e{massless}.

Feeding back \e{only} the timelike/longitudinal purely \e{potential}
part of the four-vector potential in this causal Lorentz gauge does
\e{not} affect the properties of the transverse dynamical massless
gauge particles but does transform the \e{potential field} to one
having \e{Yukawa character}.  In fact, there is no reason to believe
that such feedback would result in \e{only} the traditional exponen%
tially \e{decaying} Yukawa potential; exponentially \e{growing}
``Yukawa'' potential components should make their presence felt as
well.  One can at least \e{speculate} that such exponentially
\e{growing} feedback Yukawas might have a role in \e{quark confine%
ment}.  It turns out that the causal Lorentz gauge appears to be
compatible with \e{every possible ratio} of exponentially growing
Yukawa potential component to exponentially decaying Yukawa potential
component.  This result undoubtedly has to do with the limitations of
considering Abelian gauge theory only.  One might entertain the hope
that these ratios of exponentially growing Yukawa potential component
to exponentially decaying Yukawa potential component might eventually
become tied to non-Abelian gauge \e{color} in such a way that \e{color
singlet} states experience only the traditional exponentially \e{decay%
ing} Yukawa potential.  If the \e{confinement} of non-color-singlet
quark/gluon configurations is indeed effected by exponentially \e{grow%
ing} Yukawa potentials, it seems rather obvious that quark-gluon ``plas%
mas'' which are too ``hot'' (thermally disturbed) to readily reorganize
themselves into color singlets (``hot'' enough to be ``colorful'') will
\e{not} behave \e{at all} like a free gas.

We now proceed to the mathematical detail in Abelian gauge theory of the
results we have been outlining above in words.  We begin by considering
the consequences of the imposition of the Lorentz condition on Abelian
gauge theory, and continue to the development of the full causal Lorentz
gauge and the consequence it has of allowing the four-vector potential
to be covariantly cleaved into a timelike/longitudinal four-vector poten%
tial of purely potential field character and a remaining transverse four-%
vector potential of purely dynamical field character.

\subsection*{Isolation of the nondynamical four-vector potential
             in causal Lorentz gauge}

The Lorentz-covariant four-vector Abelian gauge field $A^\mu$,
\re{
     A^\mu(\vr, t) = (\phi(\vr, t), \vA(\vr, t)),
}{1a}
whose source is the Lorentz-covariant four-vector current density $j^\mu$,
\re{
     j^\mu(\vr, t) = (c\rho(\vr, t), \vj(\vr, t)),
}{1b}
is governed by a ``mixed bag'' of constraint and dynamical equations%
~\ct{Ka1} which are expressed in Lorentz-covariant notation as,
\re{
\bx A^\mu - \pa^\mu(\pa_\nu A^\nu)  =  j^\mu/c,
}{1c}
where the Lorentz-covariant four-vector first-derivative operators
$\pa_\mu$ and $\pa^\mu$ are given by,
\re{
     \pa_\mu = (c^{-1}\pa/\pa t, \del_\vr),
}{1d}
and,
\re{
     \pa^\mu = (c^{-1}\pa/\pa t, -\del_\vr),
}{1e}
which imply that the Lorentz-scalar contracted second-derivative
operator $\bx$ comes out to be,
\re{
    \bx  = c^{-2}\pa^2/\pa t^2 - \del^2_\vr.
}{1f}
Upon contracting both sides of Eq.~(1c) with $\pa_\mu$, we obtain
the \e{current conservation} source contraint,
\re{
    \pa_\mu j^\mu = 0.
}{1g}
It turns out that Eq.~(1c) \e{fails to uniquely determine} $A^\mu$.
It is readily verified that if the four-gradient of an \e{arbitrary}
Lorentz-scalar function $f(\vr, t)$ is added to $A^\mu$, i.e., if,
\re{
    A^\mu\rta A^\mu + \pa^\mu f,
}{1h}
then Eq.~(1c) continues to be satisfied.  We now take advantage of
this \e{gauge-invariance ambiguity}~\ct{Scw} of Eq.~(1c) to formally
\e{simplify} it by imposing on it the Lorentz-invariant \e{Lorentz
condition}~\ct{Scf},
\re{
   \pa_\nu A^\nu = 0, 
}{2a}
which results in its becoming just,
\re{
    \bx A^\mu =  j^\mu/c.
}{2b}
The Lorentz condition has \e{not}, however, \e{fully removed} the
gauge-invariance ambiguity, since we readily see that Eq.~(2b)
will still be satisfied after the gauge transformation of Eq.~(1h)
\e{provided} that the scalar gauge function $f(\vr, t)$ satisfies,
\re{
    \bx f = 0,
}{2c}
which is \e{restricted} gauge-invariance ambiguity.  We note that
the current conservation source constraint of Eq.~(1g) follows
from Eq.~(2b) upon taking account of the Lorentz condition of
Eq.~(2a).

Experience with electrostatics suggests that the \e{pure potential}
effects~\ct{Scm} which arise from $A^\mu = (\phi, \vA)$ are primarily
associated with $\phi$, whereas it is almost universally agreed that
the \e{dynamical, radiative} effects that arise from $A^\mu$ are
associated with the \e{transverse part} of $\vA$~\ct{B-S}.  But such
distinctions amongst the components of $A^\mu$ are \e{not at all}
formally apparent at this stage; in fact they are \e{missing
altogether} from Eq.~(2b).  It turns out, however, that the \e{Lorentz
condition} of Eq.~(2a) is able to nail down a \e{unique longitudinal
part} $\vA_L$ of $\vA$ with the property that $\vA_L$ is a \e{pure
homogeneous functional of the time derivative of} $\phi$, and, as a
wonderful bonus, that the four-vector $(\phi, \vA_L)$ \e{also}
satisfies the Lorentz condition!

Upon writing out the Lorentz condition of Eq.~(2a) as,
\re{
    \del\dt\vA = -\dot\phi/c,
}{3a}
we realize its extremely close formal similarity to the
Coulomb law $\del\dt\vE = \rho$ of Maxwell's equations~%
\ct{Ka1}.  The Coulomb law of course uniquely determines
the longitudinal part of $\vE$ in terms of the charge
density $\rho$.  Here we proceed to determine the
longitudinal part of $\vA$ in terms of the time
derivative of $\phi$ in exactly the same way, and obtain,
\re{
    \vA = \vA_L + \vA_T,
}{3b}
where,
\re{
    \vA_L(\vr, t) = c^{-1}\del((4\pi)^{-1}{\ty\int}d^3\vr'\,\dot\phi(\vr', t)/
                    |\vr - \vr'|),
}{3c}
and $\vA_T$ \e{must} be transverse, i.e.,
\re{
    \del\dt\vA_T = 0.
}{3d}
That Eqs.~(3b) through (3d) satisfy Eq.~(3a) is readily verified by
use of the Coulomb potential Green's function identity,
\re{
    \del_\vr^2(1/|\vr - \vr'|) = -4\pi\dl^{(3)}\,(\vr -\vr').
}{3e}
What is thus in fact verified is that,
\re{
    \del\dt\vA_L = -\dot\phi/c,
}{3f}
which implies that the four-vector field,
\re{
    A^\mu_{(0L)}\eqdf(\phi, \vA_L),
}{3g}
\e{also} satisfies the Lorentz condition,
\re{
    \pa_\mu A^\mu_{(0L)} = 0,
}{3h}
and this, together with the fact that $A^\mu$ satisfies the
the Lorentz condition (by Eq.~(2a)), implies that,
\re{
    A^\mu_{(T)}\eqdf A^\mu - A^\mu_{(0L)},
}{3i}
\e{as well} satisfies the Lorentz condition,
\re{
    \pa_\mu A^\mu_{(T)} = 0.
}{3j}
From Eqs.~(3g), (3i) and (3b), we see that,
\re{
    A^\mu_{(T)} = (0, \vA_T),
}{3k}
which shows that $A^\mu_{(T)}$ is \e{completely transverse}.
We now further note that any four-vector field which satisfies
the Lorentz condition is \e{necessarily} Lorentz-covariant.
That is because such a four-vector field contracted with the
manifestly Lorentz-covariant four-vector differential operator
$\pa_\mu$ yields \e{zero identically}, which is a \e{manifest
Lorentz scalar}.  Therefore $A^\mu_{(0L)}$ and $A^\mu_{(T)}$
are Lorentz-covariant four-vector fields.

The Lorentz condition has thus enabled us to covariantly
separate $A^\mu$ into $A^\mu_{(T)}$, which is \e{purely
transverse} and $A^\mu_{(0L)}$, which is \e{purely timelike
and longitudinal}, with its time component being $\phi$
\e{itself}, while its longitudinal part is a pure homogeneous
functional of $\dot\phi$.  Now the current density four-vector
$j^\mu = (c\rho, \vj)$ satisfies the current conservation source
constraint given by Eq.~(1g), which can be reexpressed as,
\re{
    \del\dt\vj = -\dot\rho,
}{4a}
in extremely close analogy with Eq.~(3a) for $A^\mu = (\phi, \vA)$.
Therefore we have for the current density four-vector $j^\mu$
extremely close analogs of \e{all} the results given by Eqs.~(3)
for the four-vector gauge field $A^\mu$.  We therefore simply list
the most important of these results with a minimum of comment,
\re{
    \vj = \vj_L + \vj_T,
}{4b}
where,
\re{
    \vj_L(\vr, t) = \del((4\pi)^{-1}{\ty\int}d^3\vr'\,\dot\rho(\vr', t)/
                    |\vr - \vr'|),
}{4c}
and $\vj_T$ \e{must} be transverse, i.e.,
\re{
    \del\dt\vj_T = 0.
}{4d}
In fact,
\re{
    \del\dt\vj_L = -\dot\rho,
}{4e}
which implies that the four-vector field,
\re{
    j^\mu_{(0L)}\eqdf(c\rho, \vj_L),
}{4f}
\e{also} satisfies the current conservation constraint,
\re{
    \pa_\mu j^\mu_{(0L)} = 0.
}{4g}
We of course have that,
\re{
    j^\mu_{(T)}\eqdf j^\mu - j^\mu_{(0L)},
}{4h}
\e{as well} satisfies the current conservation constraint,
\re{
    \pa_\mu j^\mu_{(T)} = 0.
}{4i}
We note that,
\re{
    j^\mu_{(T)} = (0, \vj_T),
}{4j}
which shows that $j^\mu_{(T)}$ is \e{completely transverse}.
Since $j^\mu_{(0L)}$ and $j^\mu_{(T)}$ both satisfy the
current conservation constraint, they are therefore both
Lorentz-covariant four-vector fields.

It is interesting to use Eq.~(2b) and Eqs.~(3) to determine the
equations that are satisfied $A^\mu_{(0L)}$ and $A^\mu_{(T)}$. 
From the time component of Eq.~(2b), we, of course obtain
that,
\re{
    \bx\phi = \rho,
}{5a}
or,
\re{
\ddot\phi/c^2 - \del^2\phi = \rho.
}{5b}
Thus we see that the operators which comprise $\bx$ are
$c^{-2}\pa^2/\pa t^2$ and $-\del^2$.  Now from Eqs.~(3c),
(3e) and (4c) we readily work out that we can write
$\vA_L$ and $\vj_L$ in the very convenient schematic
operator forms,
\re{
\vA_L = c^{-1}\del(-\del^2)^{-1}\dot\phi,
}{5c}
and,
\re{
\vj_L = \del(-\del^2)^{-1}\dot\rho.
}{5d}
We clearly see that the two operators which comprise the
operator $\bx$ \e{both commute} with \e{all} of the three
operators that appear in front of $\phi$ in Eq.~(5c) for
$\vA_L$.  Now the action of $\bx$ on $\phi$ is given by
Eq.~(5a).  With that and the form of Eq.~(5d), we conclude
that,
\re{
\bx\vA_L = \vj_L/c.
}{5e}
We can now combine our result of Eq.~(5e) with the forms
for $A^\mu_{(0L)}$ and $j^\mu_{(0L)}$ given by Eqs.~(3g)
and (4f) respectively, plus Eq.~(5a), to obtain that,
\re{
\bx A^\mu_{(0L)} = j^\mu_{(0L)}/c.
}{5f}
If we combine the definitions of $A^\mu_{(T)}$ and
$j^\mu_{(T)}$ that are given in Eqs.~(3i) and (4h)
with the results of Eqs.~(5f) and (2b), we also obtain,
\re{
\bx A^\mu_{(T)} = j^\mu_{(T)}/c.
}{5g}
\indent
The four-vector potential $A^\mu_{(0L)}$ is completely
determined by $\phi$ because, as we see from Eqs.~(3g)
and (5c), its time component is $\phi$ itself and its
remaining longitudinal part is a purely homogeneous
functional of $\dot\phi$.  The relation of $\phi$ to
$\rho$, however, is given by Eq.~(5a), which
leaves open the possibility that the relation of
$\phi$ to $\rho$ is \e{inhomogeneous} and \e{not fully
determined by just $\rho$ itself}.  What we face here
is simply an aspect of the \e{restricted} gauge-%
invariance ambiguity in the face of the imposition of
\e{only} the Lorentz condition, as was pointed out
in the discussion centered on Eq.~(2c).  Indeed it
is entirely clear from Eq.~(5a) that any contribution
to $\phi$ which is \e{inhomogeneous} in $\rho$ must
satisfy \e{precisely} the relation pointed out in
Eq.~(2c).  To jettison this annoying remnant of
gauge-invariance ambiguity, we need to \e{supplement}
the Lorentz condition with a \e{restriction on the
solution space of Eq.~(5a) for} $\phi$ that \e{re%
jects} any elements which are \e{inhomogeneous} in
$\rho$.  The intuitively/physically \e{most appealing}
way to achieve this is to simply assert that the rela%
tion of $\phi$ to $\rho$ is a \e{totally causal} one
in \e{both} space and time.  This \e{uniquely} pins down
the following extremely well-known and justly celebrated
homogeneous causal \e{retarded Coulomb transformation}
of $\rho$ as the desired solution of Eq.~(5a) for $\phi$,
i.e.,
\re{
\phi(\vr, t) = (4\pi)^{-1}{\ty\int}d^3\vr'\,\rho(\vr', t-c^{-1}|\vr - \vr'|)/
|\vr - \vr'|.
}{6}
With the use of Eq.~(6), a \e{particular gauge} has at long last
been precisely determined.  This gauge is as \e{close to the
Coulomb gauge} as it is \e{possible} to get \e{without} clashing
with the tenets of special relativity.  Recall that the Coulomb-%
gauge version of $\phi$ simply \e{omits} the time retardation of
the functional on the right-hand side of Eq.~(6), and thereby
manifests an \e{intantaneous response} of $\phi$ throughout the
\e{whole of space} to \e{any change} in $\rho$, which is
relativistically problematic.  The two gauges agree \e{perfectly},
however, when $\rho$ is time-independent (static).  In that case
the $\phi$ of Eq.~(6) is \e{also} time-independent, and therefore,
from Eq.~(5c), the \e{longitudinal part $\vA_L$ of $\vA$ vanishes
identically}, which is what the Coulomb gauge \e{mandates under all
circumstances} by relativistically dubious \e{fiat}.

In the gauge determined by the Lorentz condition and Eq.~(6), which we
dub the causal Lorentz gauge, \e{there is nothing remotely dynamical
about the timelike/longitudinal four-potential}
$A^\mu_{(0L)} = (\phi, \vA_L)$, because Eq.~(6) \e{completely ties}
$\phi$ to $\rho$ without \e{the least trace} of dynamical freedom, and,
of course, Eq.~(5c) \e{just as completely ties} $\vA_L$ to $\dot\phi$.
Therefore, in causal Lorentz gauge, the timelike/longitudinal four-po%
tential $A^\mu_{(0L)}$ \e{isolates the nondynamical sector of the Abel%
ian gauge theory}, and it does so in \e{relativistically compliant fash%
ion}, satisfying its own covariant Lorentz condition, Eq.~(3h), and its
own covariant ``equation of motion'', Eq.~(5f) (to which \e{only the com%
pletely causal solution} that is set out in Eqs.~(6) and (5c) is selec%
ted), a very far cry indeed from the \e{utter disregard for relativity}
inherent in the Coulomb gauge.

Thus \e{devoid} in causal Lorentz gauge of \e{any dynamical content},
the timelike/longitudinal four-potential $A^\mu_{(0L)}$ \e{cannot be
independently quantized}; in causal Lorentz gauge the \e{independently
quantizable part of the gauge theory} resides \e{entirely} in the
relativistically compliant \e{transverse, dynamical four-potential}
$A^\mu_{(T)}$, which satisfies its own covariant Lorentz condition,
Eq.~(3j), and its own covariant transverse equation of motion, Eq.~(5g).
With the dynamical transverse and the nondynamical timelike/longitudinal
sectors of the gauge theory thus cleanly and covariantly \e{separated}
in causal Lorentz gauge, the possibility of \e{discarding} one of these
sectors may be entertained.  Whether that could be called for is an em%
pirical issue: some parton-style analyses of empirical \e{hadronic} data
have suggested that \e{quarks alone} are inadequate to account for that
data, that strong participation by \e{gluons} (independently-quantized,
transverse-spin dynamical gauge particles) is in fact required.

\subsection*{Feeding back the nondynamical four-vector potential
             in causal Lorentz gauge}

A fascinating aspect of the Lorentz condition
is that it \e{precisely parallels} the four-current
conservation constraint.  Thus a gauge field which 
adheres to the Lorentz condition could conceivably
be made to contribute to its \e{own input four-cur%
rent}, but it would need to be multiplied by a fac%
tor which has dimensions of inverse length squared
in order to be four-current compatible.  For the
transverse \e{dynamical} four-potential in causal
Lorentz gauge, such a maneuver would, at least
naively, appear to endow the \e{independently-quan%
tized} transverse gauge \e{particle} with \e{mass}.
That seems uncomfortable from a physics standpoint,
however, since the inherently \e{transverse} quan%
tized gauge particle only has \e{two} spin degrees
of freedom, not the \e{three} that a spin 1 particle
with \e{mass} evidently \e{requires}.

In causal Lorentz gauge, the dynamical transverse and nondynamical
timelike/longitudinal four-potentials, however, each \e{individually}
adheres to the Lorentz condition, and each \e{also} has its \e{own}
particular \e{individually conserved} input four-current, so it is
possible for the \e{nondynamical} timelike/longitudinal four-potential
in causal Lorentz gauge to be made to contribute to \e{itself only},
and to therefore \e{not} give rise to problematic gauge-particle
\e{mass}, but to nevertheless \e{transform} its time component $\phi$
into an entity with \e{Yukawa-like} properties.  Such a \e{fed-back}
$\phi$ in causal Lorentz gauge would very likely be endowed with an
exponentially \e{growing} Yukawa-type component in \e{addition} to a
\e{traditional} exponentially \e{decaying} Yukawa component---it
obviously requires a near-miracle for an exponentially \e{growing}
Yukawa component to \e{not} develop in consequence of feedback.  It is
naturally very tempting to speculate that such an exponentially
\e{growing} fed-back $\phi$ might be the mechanism of permanent quark-%
gluon confinement.

A technical/mathematical difficulty with exponentially
\e{growing} Yukawa-type potentials is that since they
\e{cannot} be spatially Fourier transformed, \e{neither}
can they ever be the \e{result} of \e{any} approach which
\e{entails spatial Fourier analysis} for obtaining the
feedback potential response occasioned by an external
charge density.  Convenient handling of such potentials
might conceivably entail unusual techniques such as use
of the Laplace transformation.  In what follows, we cope
with this issue by first writing down a standard \e{inverse}
space-time Fourier-transformation expression which applies
to the causal retarded exponentially-\e{decaying} Yukawa
feedback homogeneous potential response to a given external
charge density, on which we carry out \e{closed-form evalua%
tion} of the \e{spatial part} of  this inverse Fourier trans%
formation in order to obtain the explicit decaying Yukawa po%
tential response \e{directly in configuration space}, with
\e{only its time} still Fourier-transformed, and then \e{demon%
strate} that this result can be straightforwardly \e{extended}
to a sizable \e{class} of causal retarded exponentially-\e{grow%
ing} Yukawa-type feedback homogeneous potential responses to the
\e{very same} external charge density.  Some physically-based
criterion for \e{choosing} amongst the \e{many different types}
of causal retarded exponentially-growing Yukawa feedback homogen%
eous potential responses to an external charge density will even%
tually need to be developed: perhaps in the more realistic non-%
Abelian color-gauge context, that choice can somehow be tied to
\e{color} in such a way that color \e{singlet} states uniquely
encounter the exponentially-\e{decaying} Yukawa feedback potential,
whilst any \e{nonsinglet} color configuration encounters an exponen%
tially-\e{growing} version of the Yukawa feedback retarded homogen%
eous potential which serves to effectively \e{confine} that nonsing%
let color configuration.

Without feedback, we recall that in causal Lorentz gauge the nondy%
namical timelike/longitudinal four-potential consists of
$(\phi, \vA_L)$, where the longitudinal three-potential $\vA_L$ is,
in fact, the homogeneous functional of $\dot\phi$ that is given by
Eq.~(5c) in such a way that the Lorentz condition $\del\dt\vA_L
+ \dot\phi/c = 0$ is \e{identically satisfied}.  Similarly, in caus%
al Lorentz gauge, this four-potential's conserved four-current source
$(c\rho, \vj_L)$ has its longitudinal part $\vj_L$ given as a homogen%
eous functional of $\dot\rho$ by Eq.~(5d) in such a way that the cur%
rent conservation constraint $\del\dt\vj_L + \dot\rho$ is \e{identical%
ly satisfied}.  The equation satisfied by $(\phi, \vA_L)$ in terms of
its source four-current $(c\rho, \vj_L)$ is,
\[\bx(\phi, \vA_L) = (c\rho, \vj_L)/c.\]
Now given a \e{nonnegative constant} $\ka$ whose dimension is inverse
length, we proceed to make the nondynamical timelike/longitudinal
$(\phi, \vA_L)$ four-potential \e{contribute to its own source} by
\e{adding} to its above \e{purely external} causal Lorentz gauge source
current the conserved ``nondynamical timelike/logitudinal four-potential
self-current'' $(-c\ka^2\phi, -c\ka^2\vA_L)$ to produce the new \e{net}
four-current source, $(c(\rho - \ka^2\phi), \vj_L - c\ka^2\vA_L)$,
which, in view of the Lorentz condition, clearly \e{also} \e{identically
satisfies} the current conservation constraint. If we now \e{replace}
the original current source $(c\rho, \vj_L)$ by this new \e{fed-back}
net current source, the equation satisfied by $(\phi, \vA_L)$ becomes,
\[\bx(\phi, \vA_L) = (\rho - \ka^2\phi, \vj_L/c - \ka^2\vA_L),\]
or,
\[(\bx + \ka^2)(\phi, \vA_L) = (\rho, \vj_L/c).\]
We in fact \e{only} need solve the \e{time-component} feedback equation,
\re{
    (\bx + \ka^2)\phi = \rho,
}{7a}
for $\phi$, because, in view of the Lorentz condition, $\vA_L$
is \e{still} the homogeneous functional of $\dot\phi$ which is
given by Eq.~(5c).  If we restrict ourselves to merely solving
for a \e{Green's function} $G(\vr, t; \ka)$ which satisfies,
\re{
    (\bx + \ka^2)G(\vr, t; \ka) = \dl(t)\dl^{(3)}\,(\vr),
}{7b}
then a solution of Eq.~(7a) which is \e{homogeneous in
the charge density} $\rho$, a \e{key property that we
insist on}, will be given by,
\re{
    \phi(\vr, t) = {\ty\int}dt'd^3\vr'\,\rho(\vr', t')G(\vr -\vr', t-t', \ka).
}{7c}
Now a causal \e{retarded} Green's function that satisfies,
\re{
    G(\vr, t; \ka) =0\ \m{when $t < 0$},
}{7d}
and whose static reduction is a \e{purely exponentially
decaying} Yukawa potential, is given by the \e{inverse
space-time Fourier transformation} expression~\ct{Scw},
\re{
    G_{-1}(\vr, t; \ka) = (2\pi)^{-4}{\ty\int_{-\infty}^{+\infty}d\om\,
    e^{-i\om t}\int}d^3\vk\,e^{i\vk\dt\vr}
    [-((\om/c) + i\ep)^2 + |\vk|^2 + \ka^2]^{-1}.
}{7e}
Because,
\re{
    (\bx + \ka^2) = c^{-2}\pa^2/\pa t^2 - \del^2 +\ka^2,
}{7f}
it is easily seen $G_{-1}(\vr, t; \ka)$ satisfies the basic Green's
function requirement of Eq.~(7b).  Since both poles in the $\om$%
-dependence of the integrand of $G_{-1}$ occur in the lower-half
$\om$-plane, and because, for $t < 0$, the $\om$ contour must be closed
in the upper-half $\om$-plane, we see that for $t < 0$, $G_{-1}$ vanish%
es, which makes it a causal \e{retarded} Green's function.  With the aid
of a contour integration, the \e{inverse spatial} $d^3\vk$-integration
of Eq.~(7e) for $G_{-1}$ can be analytically carried out in \e{closed
form}.  We have proceeded to straightforwardly \e{extend} the \e{parti%
cular} expression which thereby results for $G_{-1}$ to a \e{class} of
objects whose \e{members} we denote as $G_\et(\vr, t; \ka)$.  We will now
demonstrate that \e{each} such object $G_\et$ satisfies the basic Green's
function requirement of Eq.~(7b) \e{irrespective} of the value of $\et$,
while \e{still retaining} the causal \e{retarded} nature of $G_{-1}$.  The
expression for $G_\et$ is in the form of an \e{inverse time Fourier trans%
formation} times the factor $(2\pi)$,
\re{\ba{c}
                    G_\et(\vr, t; \ka)\eqdf                              \\
\m{}                                                                     \\
         {\ty\int_{-\infty}^{+\infty}}d\om\,e^{-i\om t}    
   [\th(1-(\om/(c\ka))^2)g_\et(r, \ka(1 - (\om/(c\ka))^2)^\hf) +
     \th(1-(c\ka/\om)^2)h(r, \om(1 - (c\ka/\om)^2)^\hf)],                \\
\m{}
\ea
}{8a}
where $r\eqdf|\vr|$, $\th$ is the standard Heaviside unit step
function, which equals zero for negative argument and unity for
positive argument, and
\re{
 g_\et(r, \ka')\eqdf(8\pi^2(r + \ep))^{-1}[\cosh(\ka'r) + \et\sinh(\ka'r)],
}{8b}
\m{}
\re{
 h(r, \om')\eqdf(8\pi^2(r + \ep))^{-1}e^{i\om'r/c},
}{8c}
and $\ep$ is a positive infinitesimal length.  The expression in
square brackets in the \e{integrand} of Eq.~(8a) equals
$(2\pi)^{-1}$ times $G_\et(\vr, \om; \ka)$, which is the \e{time
Fourier transformation of} $G_\et(\vr, t; \ka)$.  From that expres%
sion and Eq.~(8c) we can see that for \e{asymptotically large values
of} $|\om|$ (i.e., asymptotically high Fourier frequencies),
$G_\et(\vr, \om; \ka)$ behaves as $(4\pi(r + \ep))^{-1}e^{i\om r/c}$,
which is an \e{outgoing spherical wave} in light of the
time-dependence $e^{-i\om t}$ of the integrand of Eq.~(8a).  This
purely \e{outgoing} spherical wave high-Fourier-frequency asymptotic
behavior of the integrand of the inverse time Fourier transformation
of $G_\et(\vr, t; \ka)$ shows that $G_\et(\vr, t; \ka)$ is a causal
\e{retarded} Green's function.  We \e{also} note from Eqs.~(8a)
through (8c) that at the somewhat confusing ``critical points'',
$\om = \pm c\ka$, $G_\et(\vr, \om; \ka)$ is \e{continuous} as a func%
tion of $\om$ and assumes the value $(4\pi(r + \ep))^{-1})$ \e{irres%
pective} of the value of $\et$, which ensures, for \e{any} value of
$\et$, that $G_\et(\vr, \om; \ka)$ is a sensibly \e{continuous} well-%
defined function of $\om$.  It turns out that the $G_\et(\vr, t; \ka)$
of Eq.~(8a) can be expressed in terms of Bessel functions~\ct{Scw},
but it doesn't appear to be necessary or worthwhile to write them as
such here.  The positive infinitesimal length $\ep$ in Eqs.~(8b) and
(8c) can usually be dropped, but the key exception is that, if one
bears in mind the identity,
\re{
    \del^2f(r) = r^{-1}d^2(rf(r))/dr^2,
}{8d}
then meticulously careful calculation yields,
\re{
    -\del^2((8\pi^2(r + \ep))^{-1}) = (4\pi^2r)^{-1}\ep/(r + \ep)^3
    \stackrel{\ep\rta 0+}{\rta}(2\pi)^{-1}\dl^{(3)}\,(\vr).
}{8e}
From Eqs.~(8b) through (8e), it is readily shown that,
\re{
    -\del^2g_\et(r, \ka') = (2\pi)^{-1}\dl^{(3)}\,(\vr) -
    (\ka')^2g_\et(r, \ka'),
}{8f}
and,
\re{
    -\del^2h(r, \om') = (2\pi)^{-1}\dl^{(3)}\,(\vr) +
    (\om'/c)^2h(r, \om').
}{8g}
From Eqs.~(8f) and (8g), together with Eq.~(7f) we obtain,
\re{
    (\bx + \ka^2)(e^{-i\om t}g_\et(r, \ka(1 - (\om/(c\ka))^2)^\hf)) = 
    (2\pi)^{-1}e^{-i\om t}\dl^{(3)}\,(\vr),
}{8h}
and,
\re{
    (\bx + \ka^2)(e^{-i\om t}h(r, \om(1 - (c\ka/\om)^2)^\hf)) = 
    (2\pi)^{-1}e^{-i\om t}\dl^{(3)}\,(\vr),
}{8i}
and therefore,
\re{
    (\bx + \ka^2)G_\et(\vr, t; \ka) = (2\pi)^{-1}\dl^{(3)}\,(\vr)
    {\ty\int_{-\infty}^{+\infty}}d\om\,e^{-i\om t} = \dl(t)\dl^{(3)}\,(\vr),
}{8j}
which shows that \e{all} $G_\et(\vr, t; \ka)$, \e{regardless} of the
value of $\et$, are Green's functions of $(\bx + \ka^2)$.  In the
limit that we do \e{not} feed back, i.e., that $\ka\rta 0+$, we obtain,
\re{
 G_\et(\vr, t; \ka = 0) = {\ty\int_{-\infty}^{+\infty}}d\om\,e^{-i\om t}
 h(r, \om) =
 (8\pi^2r)^{-1}{\ty\int_{-\infty}^{+\infty}}d\om\,e^{-i\om(t - r/c)} =
 (4\pi r)^{-1}\dl(t - r/c),
}{9}
which is the standard causal \e{retarded} Green's function of the
Abelian gauge theory---together with Eq.~(7c) it yields the celebrated
Eq.~(6).  Note that in this no-feedback limit where $\ka\rta 0+$, there
is \e{no dependence whatsoever on} $\et$.

Finally, in the interesting case that the source charge density
$\rho(\vr, t)$ has \e{no time dependence}, i.e., is \e{static},
we readily see from Eq.~(7c) that the corresponding \e{static
Green's function,} $G_\et(\vr; \ka)$, is simply given by the
\e{integral over all time of the dynamical Green's function,}
$G_\et(\vr, t; \ka)$.  Since,
\re{
    {\ty\int_{-\infty}^{+\infty}}dt\,e^{-i\om t} = (2\pi)\dl(\om),
}{10a}
we obtain from this and Eqs.~(8a) and (8b) that,
\re{
    G_\et(\vr; \ka) = (2\pi)g_\et(r, \ka) =
                      (4\pi r)^{-1}[\cosh(\ka r) + \et\sinh(\ka r)],
}{10b}
which properly reduces to the point-charge static Coulomb
potential $(4\pi r)^{-1}$ in the limit of no feedback, i.e.,
$\ka\rta 0+$.  For $\et = -1$, it is the classic purely
exponentially \e{decaying} point-charge static Yukawa potential,
$(4\pi r)^{-1}\exp(-\ka r)$, which is the static reduction
of the classic retarded causal Fourier-transformation
Green's function that is given by Eq.~(7e).  In the case of
non-vanishing feedback, we \e{began} with this particular
inverse space-time Fourier-transformation Green's function of
Eq.~(7e), which we denoted as $G_{-1}(\vr, t; \ka)$.  We then
\e{extended} it's inverse \e{purely time} Fourier transformation
to a whole additional \e{class} of \e{extremely similar} inverse
\e{purely time} Fourier transformations for which, however,
the corresponding inverse \e{spatial} Fourier transformations
simply \e{fail to exist because those Green's functions feature
growing exponential components in configuration space}. We have
denoted the \e{members} of this class as $G_\et(\vr, t; \ka)$
for arbitrary values of $\et$, and their static reductions, which
\e{prominently display} those \e{growing} exponential components
(\e{except} for the case $\et = -1$), are given by Eq.~(10b) above.

In diametrical opposition to the classic purely exponentially
\e{decaying} static Yukawa potential associated with $\et = -1$,
there is the purely exponentially \e{growing} static ``contra-%
Yukawa'' potential associated with $\et = +1$, namely
$(4\pi r)^{-1}\exp(+\ka r)$, which although patently unavailable
via inverse spatial Fourier transformation, is \e{just as much a
legitimate consequence of the potential feedback idea} as is the
classic purely exponentially \e{decaying} Yukawa case.  Indeed the
\e{range} of $\et$-values which have \e{some admixture} of exponen%
tially {growing} static Yukawa potential component \e{utterly swamps}
the \e{single} $\et = -1$ value for which that component is precar%
iously just canceled out.  But because of the \e{immense bias} in%
troduced by the almost \e{universally employed} inverse spatial
Fourier transformation~\ct{Scw}, the \e{overwhelming prevalence} of
feedback-Yukawa static potentials which \e{grow} exponentially
\e{rather} than decay exponentially has been effectively \e{entirely
wiped off the radar screen}!

\subsection*{Conclusion}

One can at least entertain the \e{hope} that the feedback-Yukawa
exponentially \e{growing} static potential components provide a
vital clue as to the mechanism of quark-gluon confinement.  Of
course this idea, here treated only in a bare-bones Abelian gauge
theory, needs to be properly implemented in the non-Abelian
Yang-Mills color gauge theory, where the whole $\et$-range of
mixed-component exponentially growing and exponentially decaying
static Yukawa potentials can hopefully be \e{linked to color} in
such a way that \e{only the traditional} $\et = -1$ \e{purely ex%
ponentially decaying Yukawa potential} operates for \e{color singlet}
entities.  The apparently exponentially \e{growing} nature of the
confinement potential \e{also} provides some feeling for why a quark-%
gluon ``plasma'' that is too ``hot'' (strongly thermally disorganized)
to be able to readily reorganize itself into \e{unconfined color sin%
glets}---i.e., ``hot'' enough to be ``colorful''---can \e{never} have
characteristics that are \e{at all akin} to those of a free gas.

It is worthwhile to point out once again that the timelike/longitudinal
\e{potential phenomena} that we have treated here in causal Lorentz
gauge are \e{strictly non-dynamical} in nature, and thus are \e{abso%
lutely not subject to independent quantization}: the independent dynam%
ical gauge \e{quanta} are all associated to the \e{transverse} compo%
nents of the gauge field, which in causal Lorentz gauge can be cleanly
and covariantly \e{separated} from the timelike/longitudinal nondynami%
cal potential phenomena discussed here.

\end{document}